\documentclass[fleqn,10pt]{wlscirep}
\usepackage{amsmath,amssymb,graphicx}
 \usepackage{multirow}
\usepackage{xcolor}
\DeclareMathAlphabet{\mathcal}{OMS}{cmsy}{m}{n}

\title{Optical transport of sub-micron lipid vesicles along an optical nanofibre}
\author[1$\dagger$]{Takaaki Yoshino}
\author[1$\dagger$]{Daichi Yamaura}
\author[1$\dagger$]{Maki Komiya}
\author[1]{Masakazu Sugawara}
\author[1]{Yasuyoshi Mitsumori}
\author[2]{Michio Niwano}
\author[1]{Ayumi Hirano-Iwata}
\author[1]{Keiichi Edamatsu}
\author[3*]{Mark Sadgrove}
\affil[1]{Research Institute of Electrical Communication, Tohoku University, Sendai 980-8577, Japan}
\affil[2]{Kansei Fukushi Research Institute, Tohoku Fukushi University, 6-149-1 Kunimi-ga-oka, Aoba-ku, Sendai, Miyagi 989-3201, Japan}
\affil[3]{Department of Physics, Faculty of Science, Tokyo University of Science, 1-3 Kagurazaka, Shinjuku-ku, Tokyo 162-8601, Japan}
\affil[*]{mark.sadgrove@rs.tus.ac.jp}
\affil[$\dagger$]{These authors contributed equally to this work}
\date{2020/09/15}
\begin{abstract}
\textbf{Enhanced manipulation and analysis of bio-particles using light confined in nano-scale dielectric structures
has proceeded apace in the last several years. Small mode volumes, along with the 
lack of a need for bulky optical elements give advantages in sensitivity and scalability relative to conventional optical
manipulation. However, manipulation of lipid vesicles (liposomes) remains difficult, particularly in the sub-micron diameter regime. Here we demonstrate the optical trapping and transport of sub-micron diameter liposomes along an optical nanofiber using the nanofiber mode's evanescent field. We find that nanofiber diameters below a nominal diffraction limit give optimal results. Our results pave the way for integrated optical transport and analysis of liposome-like bio-particles, as well as their coupling to nano-optical resonators.}
\end{abstract}
\begin{document}
\maketitle
\section{Introduction}
The ability to confine light to sub-wavelength volumes using index contrast or plasmonic excitations has led to the rapid adoption of micro- and nano-photonics in the detection, analysis and manipulation of small particles~\cite{Nanobook, OptMan1, OptMan2,SileBubble, Aspelmeyer}. 
Of great practical interest is the use of such nanophotonic platforms for biological nanoparticles\cite{Ashkin,He,Beuwer,Sandoghdar1,Vollmer1,Small1},
where the ability to produce low-power, small-footprint, integrated optical devices can be expected to have a large impact 
in real-world applications. Among these methods, sub-wavelength diameter waveguides (often referred to as nano-waveguides in the visible regime) offer a growing platform for manipulation of bio-particles using evanescent fields~\cite{Gaugiran,Lipson}. In particular, we note reports of nano-waveguide based optical trapping and transport for micron-size cells, including red blood cells~\cite{Gaugiran, Ahluwalia, Helleso} and even live E. coli bacteria~\cite{Xin}, using a variety of waveguide configurations. Despite the growing use of plasmonics to push the boundaries of optical trapping, e.g., in relation to particle size~\cite{Pang,Hong}, nano-optical waveguides still have many advantages including in transport and delivery~\cite{Xin} and their status as a mature platform for cavity quantum electrodynamics, which may have interesting applications both fundamental  and applied for bio-nano-particles~\cite{Small1}.
\begin{figure*}[htb]
	\centering
	\includegraphics[width=0.95\linewidth]{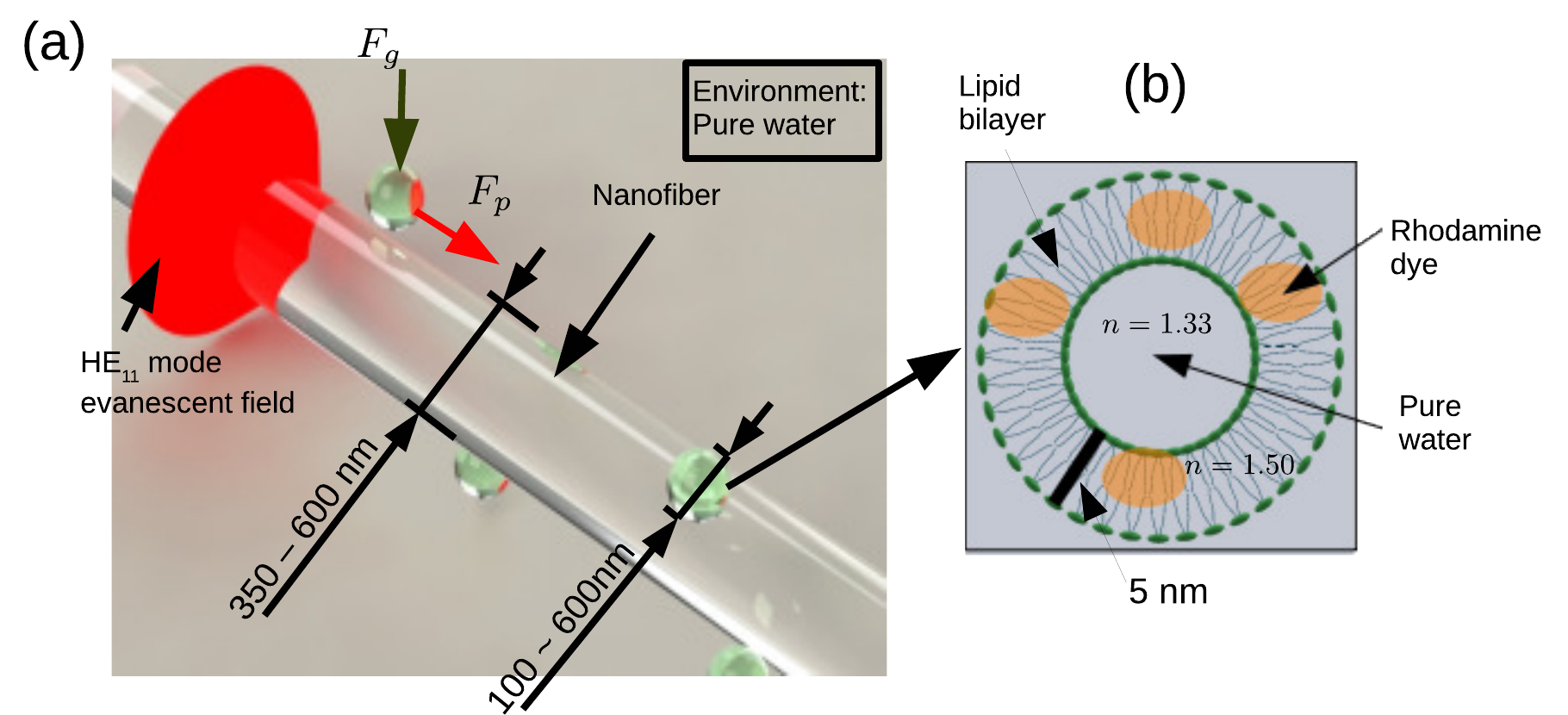}
	\caption[]{(a) Concept of the experiment. Liposomes (shown as green spheres) are trapped near the surface of the nanofiber by the evanescent field gradient force $F_g$ and propelled in the same direction as the fiber mode due to its light pressure force $F_p$.  
	(b) Schematic diagram of a unilamellar liposome tagged with rhodamine dye. 
	}
	\label{fig:Concept}
\end{figure*}

However, nano-waveguide based techniques are less advanced for lipid vesicles, also known as liposomes. These composite bio-particles consist of a lipid bilayer membrane surrounding a liquid core, and are structurally equivalent to important bio-particles including cells.
They have a number of applications including studies of cell dynamics~\cite{Somasundar,Osawa}, targeted delivery\cite{Mikhaylov,Deng}, and synthetic cell creation~\cite{Godino} to give just a few examples. Recent research has reduced the required power for optical manipulation of liposomes while extending techniques to liposome sizes below 100 nm~\cite{SizePaper,Spyratou}.  

Although some recent research has used nano-photonic methods to detect liposomes~\cite{Exo1,Exo2} and plasmonic effects have been utilized to manipulate them thermally~\cite{thermal,nanoplasmon,kaneta1,kaneta2}, no direct nano-waveguide-based manipulation of individual liposomes has been reported to our knowledge. One reason for this is that the exponential decay of the evanescent field away from the the waveguide surface along with the fact that the liposome refractive index contrast is confined to its membrane effectively means that only a small portion of the liposome experiences an optical force due to the evanescent field, making nano-waveguide based manipulation challenging.

Here, we use the evanescent field of an optical nanofiber to trap and transport liposomes with a diameter below one micron.
Compared to previous experiments which used nanowaveguides to trap and transport biological nanoparticles, our study moves the technique 
deeper into the regime of sub-micron particle size and small refractive index contrast (as small as $n=1.35$ for 200 nm diameter unilamellar liposomes used in this experiment). We also study the effect of the nanofiber diameter on the transport of liposomes and find that, counterintuitively, reducing the nanofiber diameter below the nominal diffraction limit associated with maximal evanescent field intensity produces stronger optical trapping and propulsion of liposomes. Finally, our opto-fluidic setup is very simple, requiring only a tapered fiber immersed in a droplet of solution, with no other nano- or micro-fabricated parts necessary.

\section{Experimental setup}
The situation we consider is as depicted in Fig.~\ref{fig:Concept}(a). A tapered fiber with an approximately 1 mm long waist region of diameter between 360 nm and 630 nm is immersed in a pure water solution containing liposomes. Light in the fiber's fundamental HE$_{11}$ mode creates
an evanescent field which penetrates into the surrounding solution. A particle near the fiber experiences both a gradient 
force $F_g$ which attracts the particle towards the fiber surface, and a light pressure force $F_p$ which pushes the particle in the same
direction as the nanofiber mode.
As shown in Fig.~\ref{fig:Concept}(b), we model the lipsomes as spherical lipid bilayers (index $n=1.5$~\cite{Matsuzaki}) containing pure water ($n=1.33$), with a bilayer thickness of 5 nm. The lipid membrane includes rhodamine dye, allowing the liposomes to be detected by fluorescence imaging. Fruther details about the manufacture of the liposomes used in the experiment are given in the Supplementary Material.
\begin{figure*}
	\centering
	\includegraphics[width=0.8\linewidth]{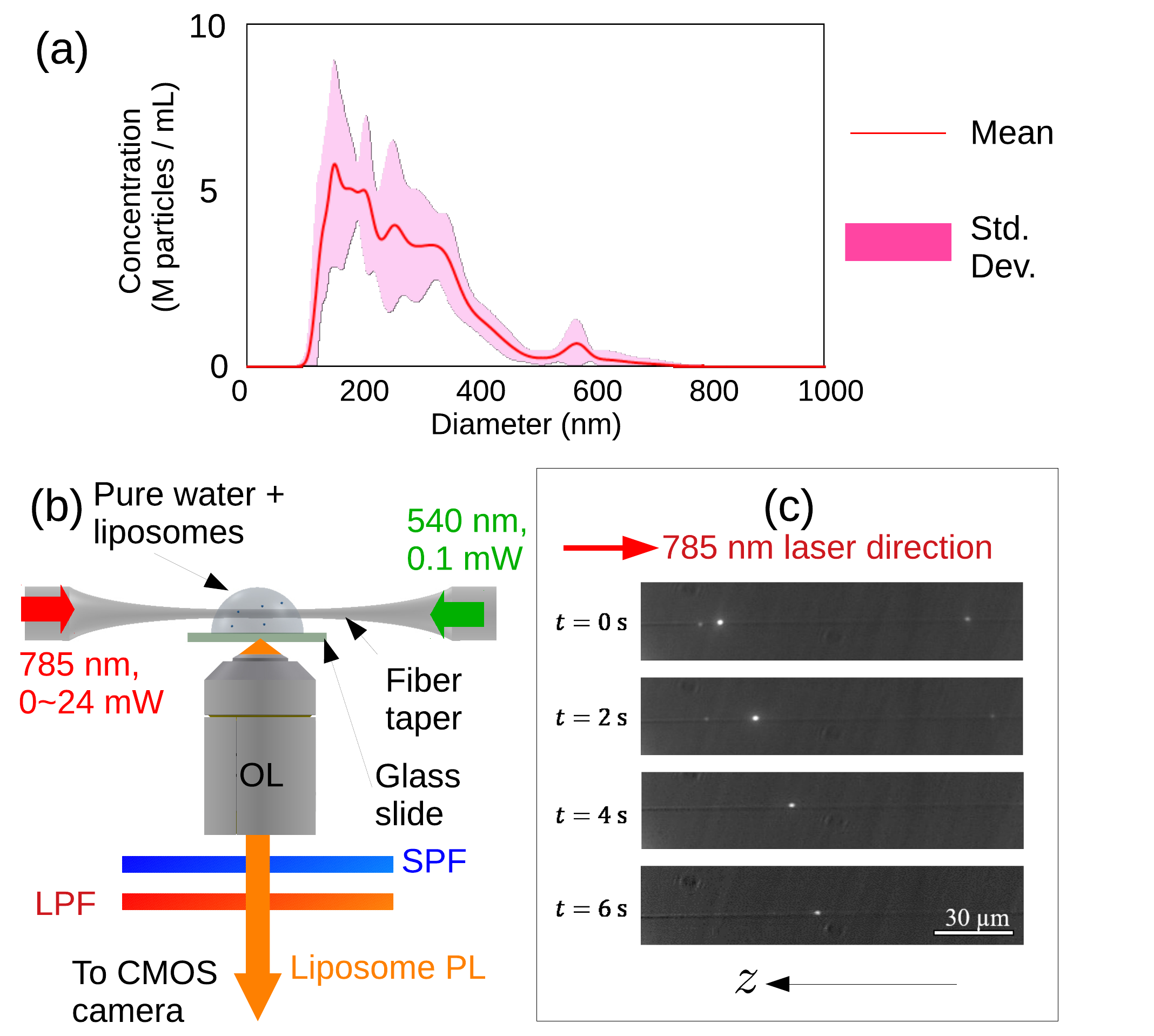}
	\caption[]{ Experimental setup. (a) Data characterizing the diameter of liposomes in the sample. 
	(b) Optical setup. A 785 nm laser injected into one end of the nanofiber provides the trapping and driving 
	force, while a 540 nm laser injected into the opposite end excites the rhodmine dye in the liposome membranes.
	Light captured by an objective lens (OL) travels through a short pass filter (SPF, cut-off wavelength 700 nm) and a long pass filter (LPF, cut-on wavelength 600 nm) leaving only the rhodamine photoluminescence (PL) which is imaged by a complementary metal oxide semiconductor (CMOS) camera. (c) Frames from a movie recorded on the CMOS camera showing the movement of a liposome along the nanofiber.}
	\label{fig:Exp}
\end{figure*}
\begin{figure*}
	\centering
	\includegraphics[width=0.75\linewidth]{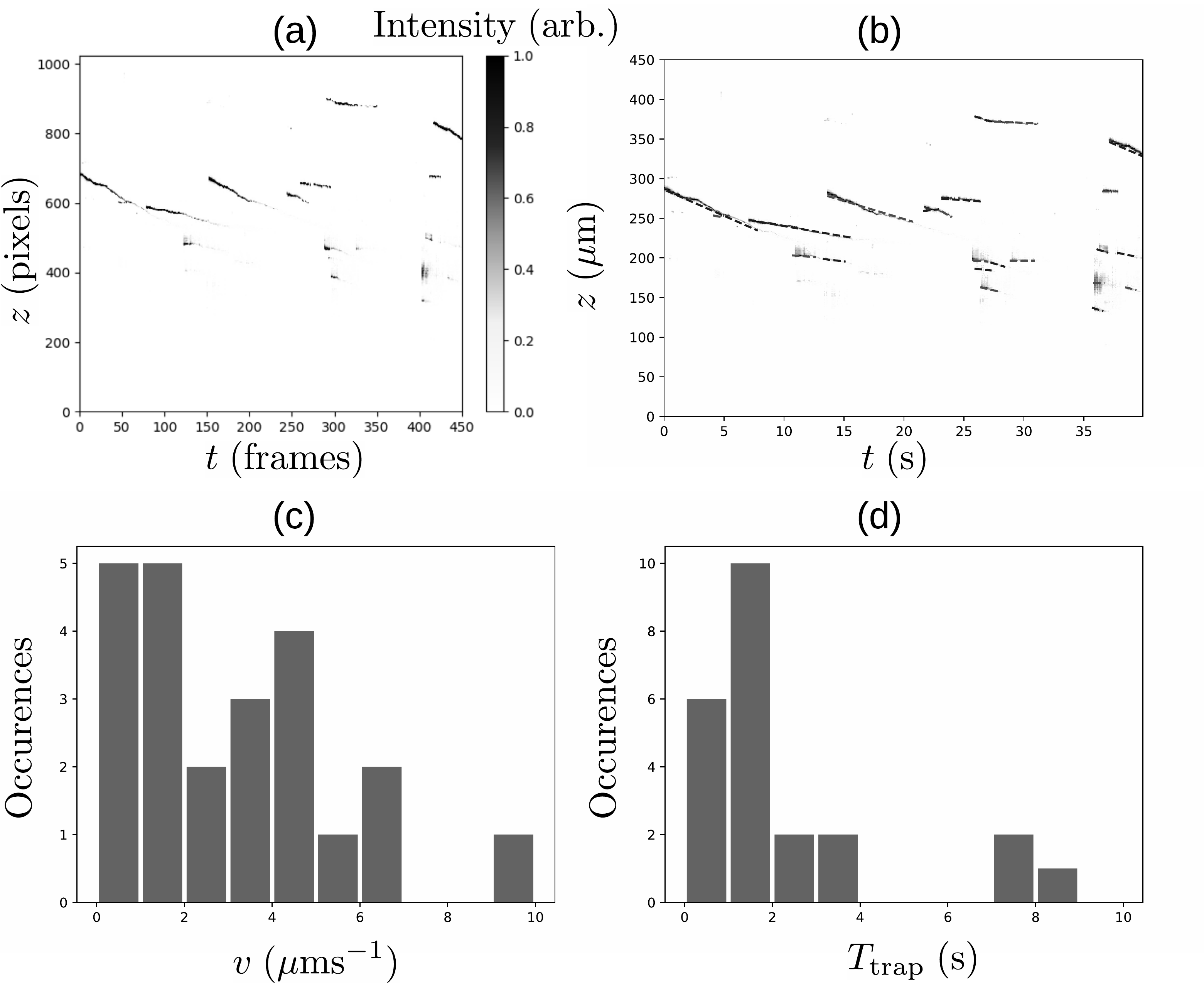}
	\caption[ ]{Observation of liposome transport along a nanofiber. (a) Raw data for a 360 nm diameter nanofiber. (b) Data
	with associated straight-line trajectories overlaid. Straight line trajectories are plotted alternately with red and blue dashed lines in order to make separate trajectories easily visible. (c) Histogram of velocities associated with each trajectory 
	in (b). (d) Histogram of trajectory lifetimes.}
	\label{fig:Results1}
\end{figure*}
Previous studies using cells found that adhesion to the nanowaveguide could prohibit cell transport along the nanowaveguide, 
and that the presence of an ionic medium,
such as that due to dissolved salt in the water surrounding the cells, was a leading contributor to this problem~\cite{Ahluwalia}. In principle, the use of surfactants may be able to alleviate this issue. However, in our current proof-of-principle experiment, we were able to avoid the problem of adhesion by using liposomes manufactured with pure water, and kept in a pure water environment during the experiment.

We characterized the size of the liposomes used in our experiments (details given in the Supplementary Material), and 
found that the distibution was as shown in Fig.~\ref{fig:Exp}(a). The liposome diameter may be seen to be distributed between 100 nm 
and 600 nm with a broad peak between 150 nm and 300 nm. The liposome density in solution was found to be $\sim 12\times10^9$ mL$^{-1}$.
It is important to note that most of the liposomes are close to or below the diffraction limit of $\lambda/2$ for the light measured in our experiment, and therefore the size of individual liposomes which are trapped and transported could not be optically determined at the time of the experiment.

The optical setup of our experiment is shown in Fig.~\ref{fig:Exp}(b). 
A nanofiber is prepared using a heat-and-pull method~\cite{SileHP} and immersed in a droplet of solution containing liposomes.  (Nanofiber diameter measurements are given in the Supplementary Material). The transmission loss due to immersion was typically less than 10$\%$.
To trap and transport liposomes near to the nanofiber surface, we introduce a 785 nm laser into the fiber. A separate laser at a wavelength of 540 nm is used to excite the rhodamine dye attached to the liposomes. The fluorescence from rhodamine has a broad spectrum with a peak near 580 nm and tails extending beyond 650 nm on the red side. We use two filters - a 600 nm cuton longpass filter and a 700 nm cutoff shortpass filter to cut the excitation and transport light leaving only the rhodamine photoluminescence from the liposomes.
We checked that in the absence of the excitation laser, nothing is observed, while in the absence of the transportation laser, liposome brownian motion near the fiber can be seen, but no transport along the fiber is observed. This confirms that the optical trapping and transport effect is due to the 785 nm fiber mode.

\section{Experimental results}
\subsection*{Trapping and transport of liposomes at the surface of a nanofiber}
Fig.~\ref{fig:Exp}(c) shows a series of images of a nanofiber (diameter 360 nm) immersed in a solution containing liposomes taken using the optical system shown in Fig.~\ref{fig:Exp}(b). The images show a liposome moving from left to right - the same direction as light propagation in the nanofiber. This data shows the trapping and transport of sub-micron liposomes by the evanescent field's gradient and scattering forces respectively. Here, we introduced white light illumination in addition to the two lasers shown in Fig.~\ref{fig:Exp}(b) to create the images, allowing the image of the nanofiber to be faintly seen. Note that we did not use white light illumination for the main experimental results reported here.
Transport is seen to occur with no sticking of the liposomes to the fiber surface.

Because of the effective one dimensional nature of transport along the nanofiber, we can combine all trajectories observed in the raw movie data into a single graph by layering the measured intensities along the fiber axis for each frame captured.
Figure~\ref{fig:Results1}(a) shows such a summarized data set for the case where the nanofiber diameter was 360 nm and the optical power 
was 24 mW. This summarized raw data was analyzed by approximating particle trajectories as straight line segments from a trajectory's start point to its end point, allowing a velocity $v$ and lifetime $T_{\rm trap}$ to be associated with each trajectory. Note that although liposomes are transported but not trapped along the nanofiber axis, because the transit time through the trapping region is typically longer than the observed trapping times, we take 
$T_{\rm trap}$ to be associated with the radial trapping potential due to the fiber mode's evanescent field.

Processed data is shown in Fig.~\ref{fig:Results1}(b), where red or blue dashed lines show the approximate straight line trajectory. (Trajectories were plotted with alternating red and blue lines to allow separate trajectories to be easily recognized). 23 trajectories were identified in the case shown.
From this analysis, the distribution of velocities and lifetimes may be extracted as shown in Figs.~\ref{fig:Results1}(c) and (d) respectively. Note that the liposomes in the sample had a wide range of diameters (150 nm - 600 nm) leading to a range of  polarizabilities and thus optical forces. This leads to the wide distribution of velocities and lifetimes that we observed. Specifically, for the case shown in Fig.~\ref{fig:Results1}, the mean velocity $\overline{v}$ was 3.1 $\mu$ms$^{-1}$ while the standard deviation was 2.4 $\mu$ms$^{-1}$. The mean trap lifetime (i.e. duration of the trajectory) was $\overline{T}_{\rm trap}=2.4$ s, while the standard deviation was 2.2 s.

\subsection*{Transport measurements for different optical powers and nanofiber diameters}
We performed similar measurements for nanofiber samples of diameter 400 nm and 550 nm. (Additional measurements at a nanofiber diameter of 630 nm produced no trajectories within the observation time frame). Measurements were made at optical powers from 0 to 24 mW at intervals of 3 mW. 
\begin{figure*}
	\centering
	\includegraphics[width=1.0\linewidth]{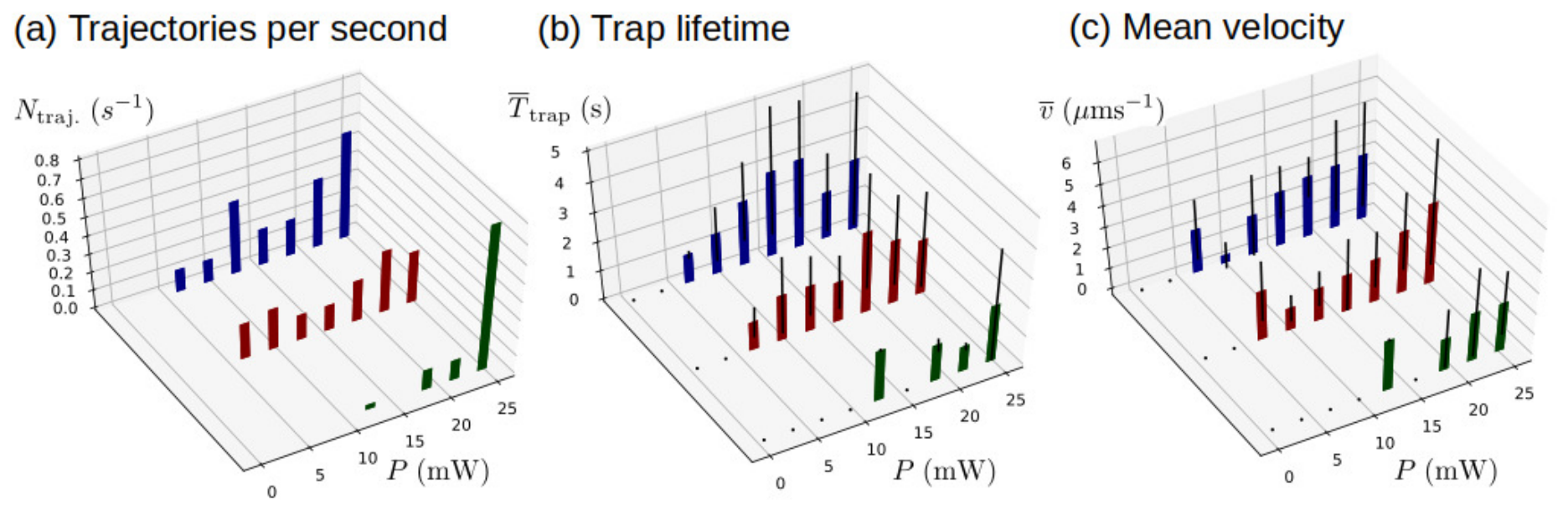}
	\caption[ ]{Power dependence of transport parameters for different nanofiber diameters. (a) Number of trajectories per second $N_{\rm traj.}$  for nanofiber diameters of 360 nm (blue bars), 400 nm (red bars) and 550 nm (green bars). 	(b) Mean lifetimes $\overline{T}_{\rm trap}$ (bars) with black lines showing $\pm 1$ standard deviation of the lifetime distribution. Nanofiber diameters are as in (a).  (c) Mean velocities $\overline{v}$ (bars) with black lines showing $\pm 1$ standard deviation of the velocity distribution. Nanofiber diameters are as in (a). }
	\label{fig:Results2}
\end{figure*}
Figure~\ref{fig:Results2} presents a summary of the results from these experiments. In each figure, blue, red and green bars correspond to 
data from the 360 nm, 400 nm, and 550 nm diameter nanofibers respectively. 

Figure~\ref{fig:Results2}(a) shows the number of trajectories per second $N_{\rm traj.}$ detected during the experiment as a function of optical power for the three different nanofiber diameters. For example, $N_{\rm traj.}$ for the case shown in Fig.~\ref{fig:Results1} was $23 / 40\;{\rm s}=0.6\;{\rm s}^{-1}$ to one decimal place. Note that in cases where ostensible single trajectories showed sections with differing velocity, we approximated the trajectory as two or more linear segments. Such a change in velocity may occur due to interaction with the nanofiber surface, or due to the combination of more than one liposome occurring during the trajectory. The number of trajectories was seen to fall as the power decreased, with a gradual fall off seen in the cases of 360 nm and 400 nm diameter nanofibers and a relatively abrupt falloff seen in the case of 550 nm diameter. It is notable that for the 360 nm and 400 nm diameter nanofibers, trapping and transport is observed for optical powers as low as 6 mW.

Figure~\ref{fig:Results2}(b) shows the mean lifetime as a function of optical power at each nanofiber diameter. Black lines show $\pm$ one standard deviation of the measured lifetimes in all cases. The lifetimes also followed the expected falloff as the optical power was decreased. However, there is a large spread in the trapping lifetimes. As discussed before, this is mainly attributable to the spread 
of liposome diameters in the sample. 

Trajectory mean velocities as a function of power are shown in Fig.~\ref{fig:Results2}(c) for each nanofiber diameter. Black lines show $\pm$ one standard deviation of the measured velocities in all cases. The data for 360 nm and 400 nm diameters are similar within the experimental error, with values between 3 and 4 $\mu$ms$^{-1}$ at 24 mW optical power, and falling below 2 $\mu$ms$^{-1}$ at 6 mW optical power. As for the lifetimes, a large spread in velocities is seen which we attribute to the variation of liposome diameter. The 550 nm diameter fiber data shows smaller velocities over all, with a mean velocity of 2.4 $\mu$ms$^{-1}$ at 24 mW, and falling to 1.6 $\mu$ms$^{-1}$ at 18 mW. Below 18 mW, only a single trajectory was seen in the case of 12 mW optical power. 

We note that both the number of trajectories and the trapping lifetimes are dependent on the optical gradient force $F_g$, whereas the velocity depends on the optical pressure force $F_p$. 

\begin{figure*}
	\centering
	\includegraphics[width=0.9\linewidth]{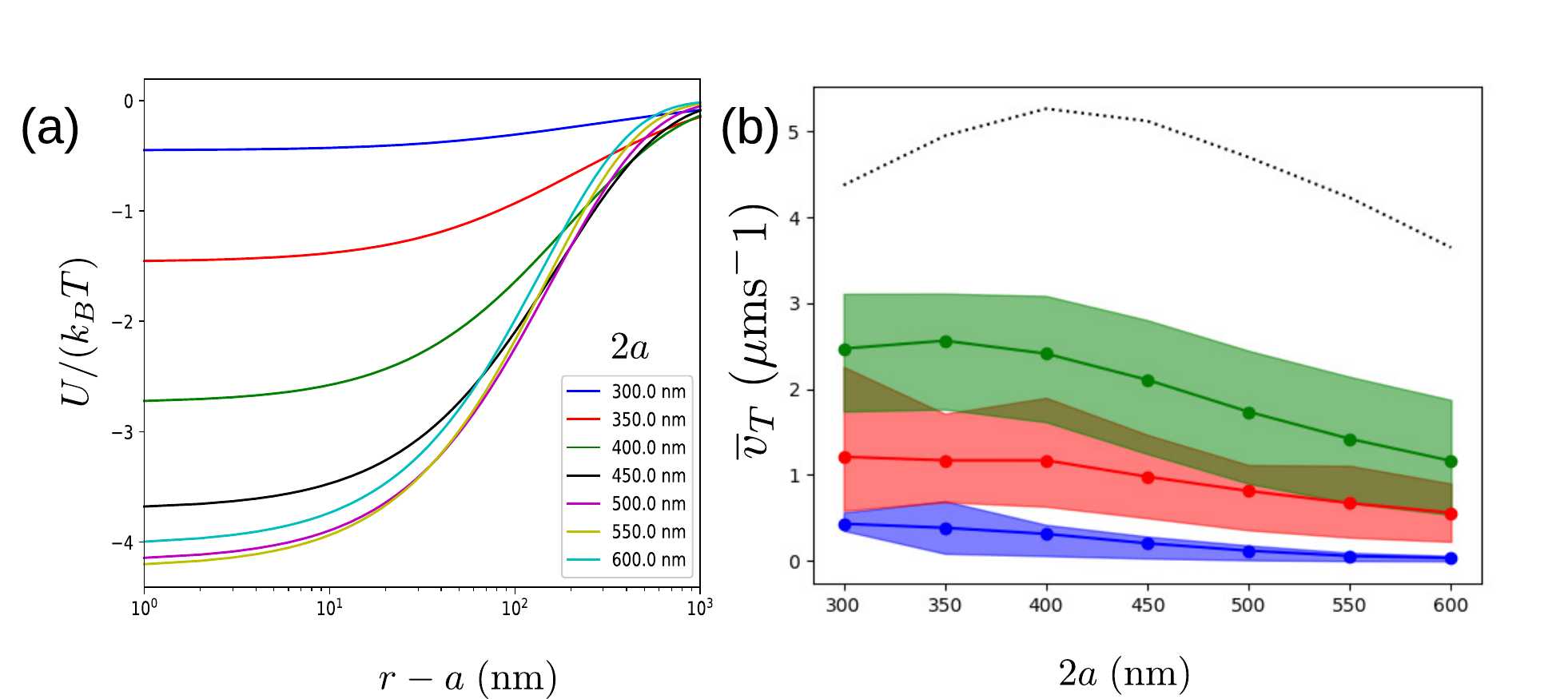}
	\caption[ ]{(a) Calculated potential energy induced by the light gradient force on a 100 nm unilamellar liposome for the nanofiber diameters $2a$ as indicated in the legend. The potential is plotted against $r-a$ so that the zero point on the horizontal axis corresponds to the nanofiber surface. The electric field intensity outside the nanofiber surface is due to a $y-$polarized HE$_{11}$ mode at 780 nm at an azimuthal angle of 90$^o$. The mode power was set to 24 mW. (b)  FDTD simulated average terminal velocities $v_T$ for unilamellar (blue line), bilamellar (red line) and trilamellar (green line) liposomes. $v_T$ was calculated for ten different liposome diameters between 100 nm and 600 nm. The shaded region in each case indicates the range of $v_T$ values from minimum to maximum. The black dotted line shows results for bilamellar liposomes when the lipid bilayer refractive index is set to 1.7.}
	\label{fig:Theory}
\end{figure*}

\section{Discussion}

The question of whether photo-thermal forces play a role in the observed trapping and transport of liposomes is important in light of recent results where liposomes were trapped using thermal gradients~\cite{thermal}.
In the highest power case considered here (24 mW of power in the fiber mode), let us assume that the 10$\%$ fall in transmission after the fiber was immersed in the liposome solution is entirely due to absorbed laser power. Then, solving the Poisson equation for temperature in a cylindrical geometry (see the Supplementary Material), and assuming that the surface of the liquid surrounding the nanofiber is at room temperature gives a rise in temperature of 10 K relative to room temperature at the nanofiber surface. 
Estimating the temperature gradient along the fiber, as an upper limit, we can assume the temperature difference between the two ends of the nanofiber is 10 K, giving a temperature gradient of $\sim10^4$ Km$^{-1}$. This is three orders of magnitude smaller than the values reported recently by Hill \textit{et al.}~\cite{thermal} in which thermal trapping of liposomes was achieved. This suggests that thermal gradient based forces are not relevant to the transport part of our current experiment. On the other hand, a temperature rise of 10 K would reduce the water viscosity in the vicinity of the fiber surface by $\sim 12\%$ compared to room temperature. This leads to a similar percentage increase in the liposome velocity relative to room temperature.
As for the radial temperature gradient, it can be estimated to be of order $1\times10^6$ Km$^{-1}$ (see the Supplementary Material), still an order of magnitude smaller than the largest gradients used by Hill \textit{et al.}~\cite{thermal} to thermally trap liposomes. Although this result does not rule out the possibility that thermal effects could play some role in trapping particles at the nanofiber surface, as we will show below, optical forces alone are sufficient to account for the observed behavior.

We now compare our experimental results with calculations of the optical force on liposomes near the nanofiber. Figure~\ref{fig:Theory}(a) shows the optical potential induced by the gradient force $F_g$ of the nanofiber mode for a number of nanofiber diameters. Calculations were performed for a liposome of 100 nm diameter (at the edge of the Rayleigh regime) assuming a spherical liposome with a uniform index of 1.37, as calculated using a volume average of the outer membrane and the enclosed water and for 24 mW of optical power. For nanofiber diameters above 300 nm, the potential depth is seen to be comparable to or greater to $k_B T$ even for this small liposome diameter, and the depth of the potential ideally increases with the liposome radius, implying that radial trapping of the liposomes in our sample can be readily achieved in agreement with our experimental results.

The optical radiation pressure force $F_P$ on the trapped liposomes is most accurately calculated by numerical simulations of Maxwell's equations. We used the finite-difference time-domain method (FDTD) to perform the force calculations. (Details are given in the Supplementary Material).  Although the liposomes used in our experiment are predominantly unilamellar, it is expected that both bilamellar and trilamellar liposomes are present in the solution. Recently, using the same preparation method as that used here, Nele \textit{et al.} reported percentages by number of 79, 14 and 6$\%$ for unilamellar, bilamellar and trilamellar liposomes respectively~\cite{Nele}, and a similar percentage composition is expected in our case. We therefore performed simulations for uni-, bi-, and trilamellar liposomes.

Assuming constant liposome velocity (i.e. terminal velocity in water), and neglecting the effect of the nanofiber, the light pressure force can be converted into a velocity using Stokes law $F_P = 3\pi\Phi_l\eta v_T$, where $\Phi_l$ is the liposome diameter, $\eta = 8.9\times10^{-4}$ Pa$\cdot$s is the dynamic viscosity of water at room temperature, and $v_T$ is the liposome (terminal) velocity.
(Corrections to Stokes law for particles moving near to a surface have been made in the case of micro-fiber based particle transport~\cite{Sile}, but it is not clear that such corrections, developed for a plane wall~\cite{Goldman}, are valid in our case. In any case, comparison with experimental results provides no evidence for a drag exceeding Stokes law).
Following our estimation of the temperature near the nanofiber surface, we adjust the velocities calculated at room temperature up by 12$\%$ to allow for the expected drop in viscosity. Figure~\ref{fig:Theory}(b) shows numerical calculations of the velocity in the case of unilamellar (blue line), bilamellar (red line) and trilamellar (green line) liposomes for nanofiber diameter $2a$ between 100 and 600 nm. The displayed values $\overline{v}_T$ are averages over simulation results for liposome diameters between 100 and 600 nm, and the shaded region in each case indicates the range of $v_T$ values from minimum to maximum. It is notable that the velocity increases as the nanofiber diameter decreases, with maximum velocities of approximately $0.4$, $1.2$ and $2.4\;\mu$ms$^{-1}$ for uni-, bi- and trilamellar liposomes respectively. In addition, the effect of rhodamine dye on the lipid index at 780 nm is unknown, but is expected to increase the index to a value between 1.5 and 2.0~\cite{Alnayli}. For comparison, we show the case for bilamellar liposomes with a membrane index of 1.7 as a dotted black line in Figure~\ref{fig:Theory}(b).
We note the trend in mean velocity as nanofiber diameter varies is in agreement with that observed in experiments. 
In particular the velocities seen in the case of 360 nm and 400 nm nanofiber diameters are higher on average than those seen
in the case of 550 nm diameter. This is somewhat surprising, given that the nanofiber diameter which optimizes the 
surface intensity is 540 nm for 785 nm input light. Relative to the intensity of the evanescent field, this value may be considered to be the diffraction limit for the nanofiber mode.

The fact that optimal transport occurs for nanofiber diameters less than this limit, even though the surface intensity drops from its maximum value, suggests that \emph{mode penetration into the surrounding solution is more important in determining the total light pressure force on liposomes than the maximum intensity at the nanofiber surface}. Although a precise study of this mechanism is still underway, intuitively, the shell structure of the liposome, where the index contrast is concentrated in a thin region, means that a higher field intensity at the far surface of the liposome (i.e. due greater mode penetration at small fiber diameter) will tend to increase the 
overall optical force experienced by the liposome.

In terms of quantitative agreement, we note that the maximum mean velocity predicted by simulations for trilamellar liposomes (2.4 $\mu$ms$^{-1}$) is close to the largest mean velocities seen in experiments ($3\sim4\;\mu$ms$^{-1}$). The larger velocities in the case of experiments may be due to the increase of the membrane refractive index due to the presence of rhodamine, and to convection within the 
liquid sample due to the temperature gradient along the nanofiber axis.

\section{Conclusion}

The results presented above clearly demonstrate the optically induced trapping and transport of liposomes along the surface of 
an optical nanofiber. This transport was achieved for moderate laser powers compared to liposome trapping using conventional 
optical tweezer techniques~\cite{SizePaper}, with trapping and transport observed down to 6 mW optical power for the two thinnest nanofiber samples.  
Our experiment can be considered to be a proof of principle. For applications, where a saline buffer may be necessary, the use of a surfactant rather than a pure water environment is possible, and an increase of power, along with introduction of a counter-propagating fiber mode can be employed to trap liposomes stably (unless transport such as that demonstrated here is desired for delivery applications).

Our results build on the studies mentioned in the Introduction where nanowaveguides have been used to trap and propel micro-sized bioparticles, and extend them to the sub-micron regime, and the limit of very low index contrast. One obvious extension of the present results would be to the trapping and delivery of exosomes which are typically below 100 nm in size.
Another intriguing avenue for future research is the use of optical cavities to induce strong light-matter coupling between bio-particles and cavity fields~\cite{Small1}. Optical nanofibers are a mature platform for nanophotonics, including photonic crystal cavities~\cite{ChandraQED,KaliReview}, making the present experiment an ideal base on which to build towards such studies.

\section*{Funding}
JSPS KAKENHI (Grant no. JP17H05460) in Scientific Research on Inovative Areas ``Nano-material optical-manipulation".

\section{Author contributions}
TY, DY, and MK contributed equally to this work.

\section*{Acknowledgments}
T. Y. acknowledges support from The Tanaka Kikinzoku Memorial Foundation scholarship, and The Ushio Foundation scholarship.
M. S. acknowledges support from JSPS KAKENHI (Grant no. JP17H05460) in Scientific Research on Inovative Areas ``Nano-material optical-manipulation".
M.S. is grateful to T. Kaneta for stimulating discussions regarding the optical manipulation of liposomes.
We thank T. Ma for helping us to measure the size distribution of the liposomes.
This research was partially performed using the facilities of the Fundamental Technology Center, Research Institute of Electrical Communication, Tohoku University.

\section*{Disclosures}
The authors declare no conflicts of interest.

\section*{Acknowledgements}
T. Y. acknowledges support from The Tanaka Kikinzoku Memorial Foundation scholarship, and The Ushio Foundation scholarship.
M. S. acknowledges support from JSPS KAKENHI (Grant no. JP17H05460) in Scientific Research on Inovative Areas ``Nano-material optical-manipulation" and is grateful to T. Kaneta for stimulating discussions regarding the optical manipulation of liposomes.
We thank T. Ma for helping us to measure the size distribution of the liposomes.
This research was partially performed using the facilities of the Fundamental Technology Center, Research Institute of Electrical Communication, Tohoku University.

\section*{Author contributions}
TY manufactured nanofibers, set up part of the experiment, performed the optical experiments, and analysed the data. DY and MK prepared liposome samples. M. Sugawara contributed to nanofiber based parts of the experiment. YM contributed to the optical part of the experiment and co-supervised the research. MN performed mesurements on the size and concentration of the liposomes. M. Sadgrove, AH and KE conceived of the original experiment and co-supervised the experiment. M. Sadgrove set up part of the experiment, performed preliminary measurements, performed simulations, and contributed to the analysis of the results. All authors discussed and contributed to writing the manuscript. 
\section*{Additional information}
Correspondence regarding the paper should be addressed to Mark Sadgrove.

\section*{Supplementary material}
\section*{Research Methods}

\subsection*{Fiber preparation and insertion into liposome solution}
Tapered fibers were fabricated using a heat and pull process~\cite{BirksLi,SileHP} applied to commercial single mode optical fiber (780HP). 
The nanofiber was submerged in a $\sim$100 $\mu$L droplet of liposome solution (liposome density $\sim 12\times10^9$ mL$^-1$) by touching the droplet to the fiber from below and then ``sinking" the nanofiber in the droplet by pipetting a small amount ($\sim$10 $\mu$L) of pure water from above.
\subsection*{Optical measurement}
We used a 50x objective (Nikon, Lu Plan Fluor 50x / NA 0.8) for observation of the nanofiber waist. The light from the lens was detected by a CMOS camera (Thorlabs DCC1545M).
 The liposome membrane was tagged with rhodamine dye whose optical absorption band exists near 550 nm, 
with a broad emission band centered near 600 nm. To excite the rhodamine and detect liposomes, we introduced a CW laser (wavelength 530 nm)
into the nanofiber. To filter out the excitation light and any scattered light from the transport beam, we inserted two filters (600 nm long pass and 700 nm short pass) before the CMOS camera. For each optical power, we saved between 35 and 40 s of data as an avi format movie file.
\subsection*{Data processing}
To process the data, we first extracted each frame from the raw movie data to separate image files. 
Secondly, we created one dimensional (1D) data from each image file by extracting pixel values along the line coinciding with the fiber position in the image. We then agglomerated the 1D data into a single 2D data set which allowed the visualization and analysis of trajectories taken by individual liposomes.  We then identified the start and end point of each trajectory and calculated the associated velocity and lifetime. In rare cases where an apparent single trajectory has sections with varying velocity, we approximated the trajectory by straight line segments, so that Stokes law could be applied to each segment.
\subsection*{Liposome sample preparation}
4 mg of 1,2-dioleoyl-sn-glycero-3-phosphocholine (DOPC) (Avanti Polar Lipids, Inc.) and 0.006 mg of 1,2-dioleoyl-sn-glycero-3-phosphoethanolamine-N-(lissamine rhodamine Bsulfonyl) (Rhod-DOPE) (Avanti Polar Lipids, Inc.) were dissolved in chloroform in a glass vial. Chloroform was evaporated with a flow of nitrogen gas for 5 min and then evacuated in a vacuum desiccator for at least 6 h. The dried lipid films were rehydrated in 5 mL of pure water to give a lipid concentration of approximately 1 mM and vortexed for 1 h to prepare phospholipid multilamellar liposome suspensions. The liposome suspension was then repeatedly frozen and thawed five times with liquid nitrogen, and extruded five times through a polycarbonate membrane filter with a pore size of 400 nm (Merck Millipore Ltd.)\cite{HiranoLab1,MacDonald}.
\subsection*{Liposome sample characterization}
The size distribution of the prepared liposomes was measured using a nanoparticle characterization system equipped with a 405 nm wavelength laser (NanoSight LM10-HSBFT14, Quantum Design Japan). The radius $r$ of liposomes, was derived using the Stokes-Einstein equation $D_t=k_B T / 6\pi\eta r$ where $k_B$ is Boltzmann's constant, $T$ is the temperature, $\eta$ is the viscosity of the liquid, and $D_t$ is the diffusion coefficient. The value of $D_t$ can be obtained based on the analysis of the Brownian motion of nanoparticles by the NTA technique. Measurements were carried out at room temperature in a cell installed in the NTA equipment.

\section*{Finite difference time domain simulations}

Here, we provide more details regarding the numerical simulations of the light pressure force imposed by the fiber mode on the liposomes. This is the force which causes the liposomes to move along the nanofiber in the same direction as the mode propagation. The nature of 
the liposomes brings a number of challenges when performing simulations. In particular, we note the following: i) the low index contrast between the liposomes and the surrounding water means that the preferred (and memory efficient) method of calculation - the Maxwell stress tensor - is not suitable, ii) the lamellarity of individually trapped and transported liposomes is unknown and iii) the rhodamine introduced to the lipid layer causes an unknown increase in the liposome refractive index. We now consider these points in order:

We performed numerical evaluation of the optical forces on liposomes near the nanofiber surface by evaluating 
the Lorentz force throughout a 1 $\mu$m$^3$ volume which included the liposome but excluded the nanofiber. This "volumetric" method
is more accurate than the usual Maxwell stress tensor method for particles which have small index contrast with the surrounding medium.
Specifically, for particles of very small contrast with the surrounding medium, numerical calculation of the Maxwell stress tensor
corresponds to finding a very small difference between two large numbers, and is therefore prone to numerical noise~\cite{Lumerical}.
\begin{figure}[htb]
	\centering
	\includegraphics[width=0.9\linewidth]{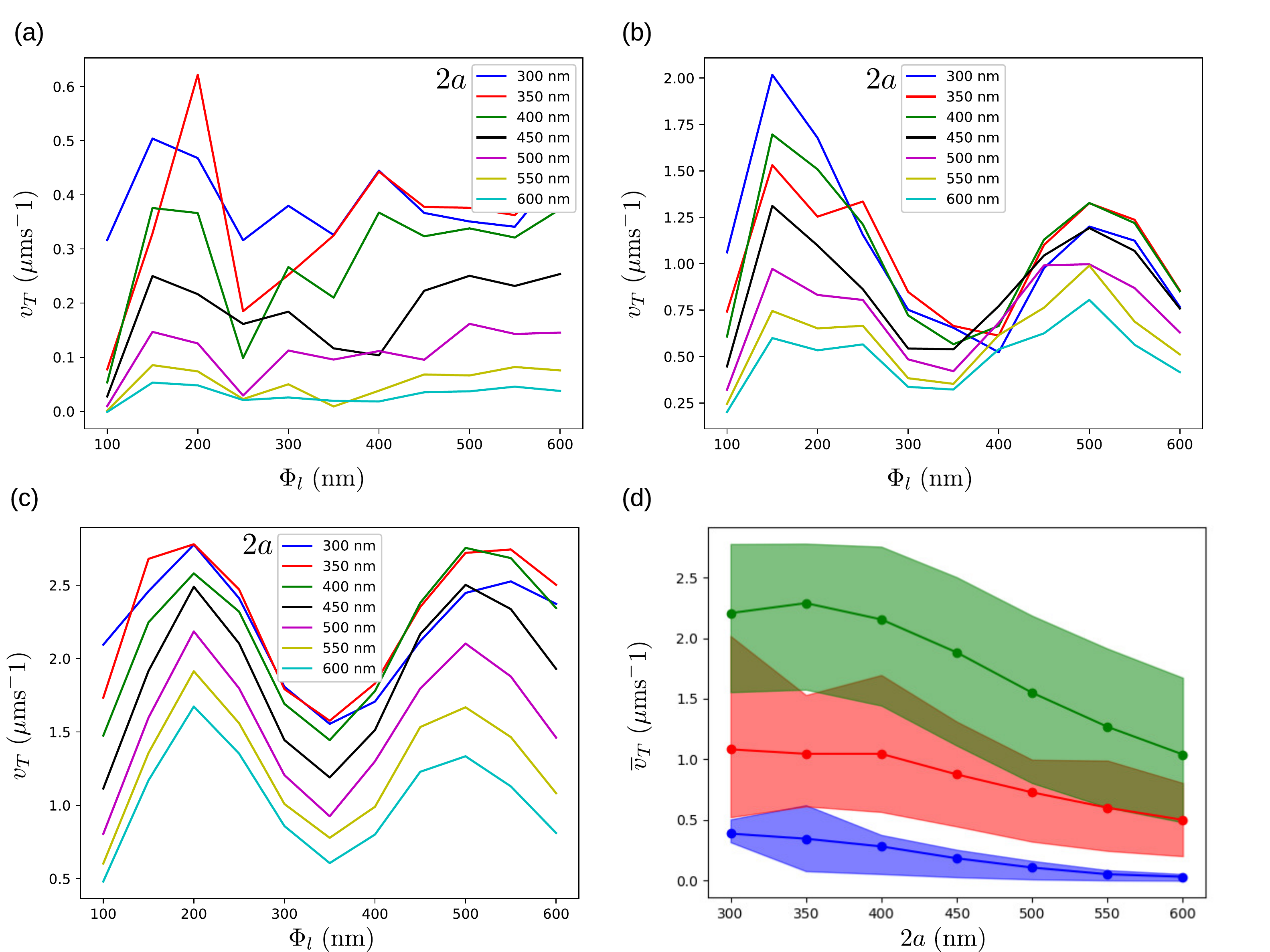}
	\caption[]{(a) Numerically calculated terminal velocities as a function of unilamellar liposome diameter for various nanofiber diameters as indicated in the legend. (b) Same as (a) but for bilamellar liposomes. (c) Same as (a) but for trilamellar liposomes. 
	(d)	Average velocities over liposome diameter as a function of nanofiber diameter $2a$, for unilamellar (blue circles), bilamellar (red circles) and trilamellar (green circles) liposomes. Lines are included to guide the eye. The shaded region in each case indicates the range of $v_T$ values from minimum to maximum. }
	\label{fig:FDTD}
\end{figure}
Second, we model the liposomes as 5 nm shells of dielectric material of 
index $n_l=1.5$~\cite{Matsuzaki} encasing pure water ($n_w=1.33$). 
Although the liposomes are predominantly unilamellar, it is expected that both bilamellar and trilamellar liposomes are present in the solution. Recently, using the same preparation method as that used here, Nele \textit{et al.} reported percentages by number of 79, 14 and 6$\%$ for unilamellar, bilamellar and trilamellar liposomes respectively~\cite{Nele}, and a similar percentage composition is expected in our case.

Third, the liposomes are tagged with rhodamine-B dye to allow their detection by photoluminescence measurements. The addition of
rhodamine dye is expected to cause an increase in refractive index of the lipid membrane. In particular, Alnayli et al. have reported 
values of refractive index above 710 nm for Rhodamine B in solution of between 1.5 and 2.0 depending on concentration~\cite{Alnayli}. Although an exact measurement of the refractive index of the lipid membrane of the liposomes used in our experiment is beyond the scope of the present work, it is necessary to consider the possible range of refractive index when comparing experiment and simulations.

The numerical simulations of the light pressure force $F_P$ on the liposomes were performed using the finite difference time domain method (Lumerical FDTD). We converted the force into a velocity by assuming that the light pressure force was balanced by the viscous force of the water, and using Stokes formula $F_P = 3\pi\Phi_L\eta v_T$, where $\Phi_L$ is the liposome diameter, $\eta = 8.9\times10^{-4}$ Pa$\cdot$s is the dynamic viscosity of water at room temperature and $v_T$ is the liposome (terminal) velocity. (Note that in the main paper, we adjusted the velocities to allow for the 10 K increase in temperature expected at the fiber surface, which leads to a $\sim 12\%$ decrease in viscosity relative to room temperature, and an associated $\sim 12\%$ increase in the calculated velocity).
The use of the unmodified Stokes formula may produce an overestimate of the liposome velocity given that the presence
of the nanofiber should lead to additional resistance to the liposome motion. Corrections to Stokes law due to the presence of 
a plane surface which have been applied in experiments using microfibers are not obviously applicable here. More importantly, we find no experimental evidence to suggest that the true drag is significantly larger than that predicted by Stokes law. A more quantitative comparison of experiment and theory in this regard is beyond the scope of the present study. 

The velocity is shown as a function of liposome diameter for unilamellar liposomes in Fig.~\ref{fig:FDTD}(a) for various nanofiber diameters (NFDs) as indicated in the legend. 
We see that the mean velocity increases as the nanofiber diameter is decreased from 600 nm to 300 nm. The maximum velocity of $\approx 0.6$ $\mu$m$s^{-1}$ is seen to occur when the fiber diameter is 350 nm, and the liposome diameter is 200 nm. In Fig.~\ref{fig:FDTD}(b), the same results are shown for the case where the liposome is bilamellar (i.e. the lipid membrane is twice as thick). Roughly the same trend is seen with increasing velocity as the nanofiber diameter is decreased from 600 nm to 300 nm. However, the average value of the velocity is
approximately three times that see for unilamellar liposomes. The maximum value of about 2 $\mu$m$s^{-1}$ occurs for a nanofiber diameter of 300 nm and a liposome diameter of 150 nm. Additionally, two broad peaks in the velocity are seen near values of $\Phi_L=200$ nm
and $\Phi_L=500$ nm, suggesting weak resonant effects dependent on size (i.e. whispering gallery modes). In Fig.~\ref{fig:FDTD}(c) shows the same calculations in the case of trilamellar liposomes. Here we see average velocities about 5 times higher than the unilamellar cases with a maximum velocity of about 2.7 $\mu$ m$s^{-1}$ seen for $\Phi_d=200$ nm and $500$ nm, and for a fiber diameter of 350 nm. Finally, In Fig.~\ref{fig:FDTD}(d) We show a summary of the data in (a)-(c) by plotting mean velocities over liposome diameter as a function of the nanofiber diameter $2a$.

\section*{Estimation of temperature at nanofiber surface}
\begin{figure}[htb]
	\centering
	\includegraphics[width=0.5\linewidth]{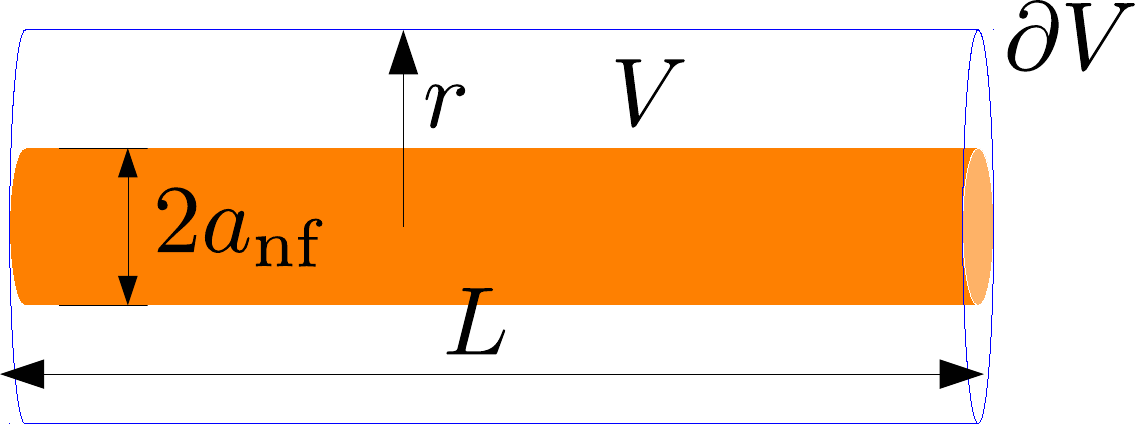}
	\caption[]{Schematic diagram of the system we consider in order to evaluate the temperature rise due to absorbance 
	of laser light from the fiber mode. A region of diameter $a_{\rm nf}$ and length $L$ is considered to have a constant 
	power density $\rho$. We solve the Poisson equation for a cylindrical volume $V$ with surface $\partial V$ outside the constant power region with radius $r>a_{\rm nf}$.  }
	\label{fig:Poisson}
\end{figure}
We estimated the temperature at the nanofiber surface and the temperature gradient experienced by liposomes using a simplified model
of a cylindrically symmetric heat source (light absorbed from the guided mode of the nanofiber) as shown in Fig.~\ref{fig:Poisson}.
Assuming an infinitely long nanofiber (in our case, since the nanofiber length is much longer than its radius, this is not a bad approximation), the Poisson equation for the temperature $T$ is
\begin{equation}
\nabla^2 T=\left\{\begin{array}{l}
-\frac{\rho}{\lambda},\;r<a_{\rm nf} \\
0,\;r>a_{\rm nf}
\end{array}\right.,
\end{equation}
where $\rho=2.4\;{\rm mW} / (\pi\times (200\;{\rm nm})^2\times 1\;{\rm mm})$ is the power density defined as the total absorbed power within the region of radius $a_{\rm nf}$ and over the 1 mm region
where the nanofiber is at its thinnest, and $\lambda = 0.592$ W/m.K is the thermal conductivity of water.

It is simplest to first use Gauss' law to solve for the gradient of the temperature:
\begin{eqnarray}
\oint_{\partial V} \nabla T\cdot d\textbf{A} & = &  \int_V -\frac{\rho}{\lambda}{\rm d}V\nonumber\\
\Rightarrow 2\pi r L (\nabla T)_r & = & -2\pi\frac{\rho a_{\rm nf}^2 L}{\lambda}\nonumber\\
\Rightarrow \frac{\partial }{\partial r}T & = & -\frac{\rho a_{\rm nf}^2}{\lambda r}.\label{eq:Gauss}
\end{eqnarray}

Integrating Eq.~\ref{eq:Gauss} gives a general solution for $T$:
\begin{equation}
T(r) = -\frac{\rho a^2_{\rm nf}}{\lambda}\log(r) + C.
\end{equation}

The constant $C$ can be determined by imposing an appropriate boundary condition. Here, we will assume that at the surface of the water droplet, a radial distance $R$ from the fiber center, the water is at room temperature $T_0$. This leads to the equation
\begin{equation}
C = T_0 + \frac{\rho a^2_{\rm nf}}{\lambda}\log(R).
\end{equation}

Finally, we find that the temperature difference $\Delta T$ between the fiber surface and room temperature can be written
\begin{equation}
\Delta T(r) = \frac{\rho a^2_{\rm nf}}{\lambda}\log\left(\frac{R}{r}\right).
\end{equation}

Taking $a_{\rm nf}$ to be 200 nm, and $R$ to be 1 mm, we find that $\Delta T\approx 10$ K. This value is insensitive to the exact values of 
$a_{\rm nf}$ and $R$, being sensitive instead to the difference in their orders of magnitude due to the logarithmic dependence.

The temperature gradient in the radial direction can now be calculated. As a representative example, we set the nanofiber radius to $a_{\rm nf}=200$ nm, and choose a liposome diameter $\Phi_l=300$ nm. Then, using Eq.~\ref{eq:Gauss}, the temperature gradient at the center of the 
liposome is found to be about $3\times 10^6$ Km$^{-1}$. 

The gradient along the nanofiber is more difficult to model accurately. As an upper bound, we can assume the the entire temperature gradient $\Delta T$ occurs along the $\sim 1$ mm length of the nanofiber. Then, in a one dimensional approximation, the temperature gradient is $10 {\rm K} / 1\;{\rm mm}=10^4$ Km$^{-1}$. 

\section*{Nanofiber diameter measurements}
Here, we show typical results for diameter measurements of optical nanofibers. The measurements were made after the experiment 
was finished and the nanofibers had been removed from the liposome solution.
\begin{figure}[htb]
	\centering
	\includegraphics[width=0.6\linewidth]{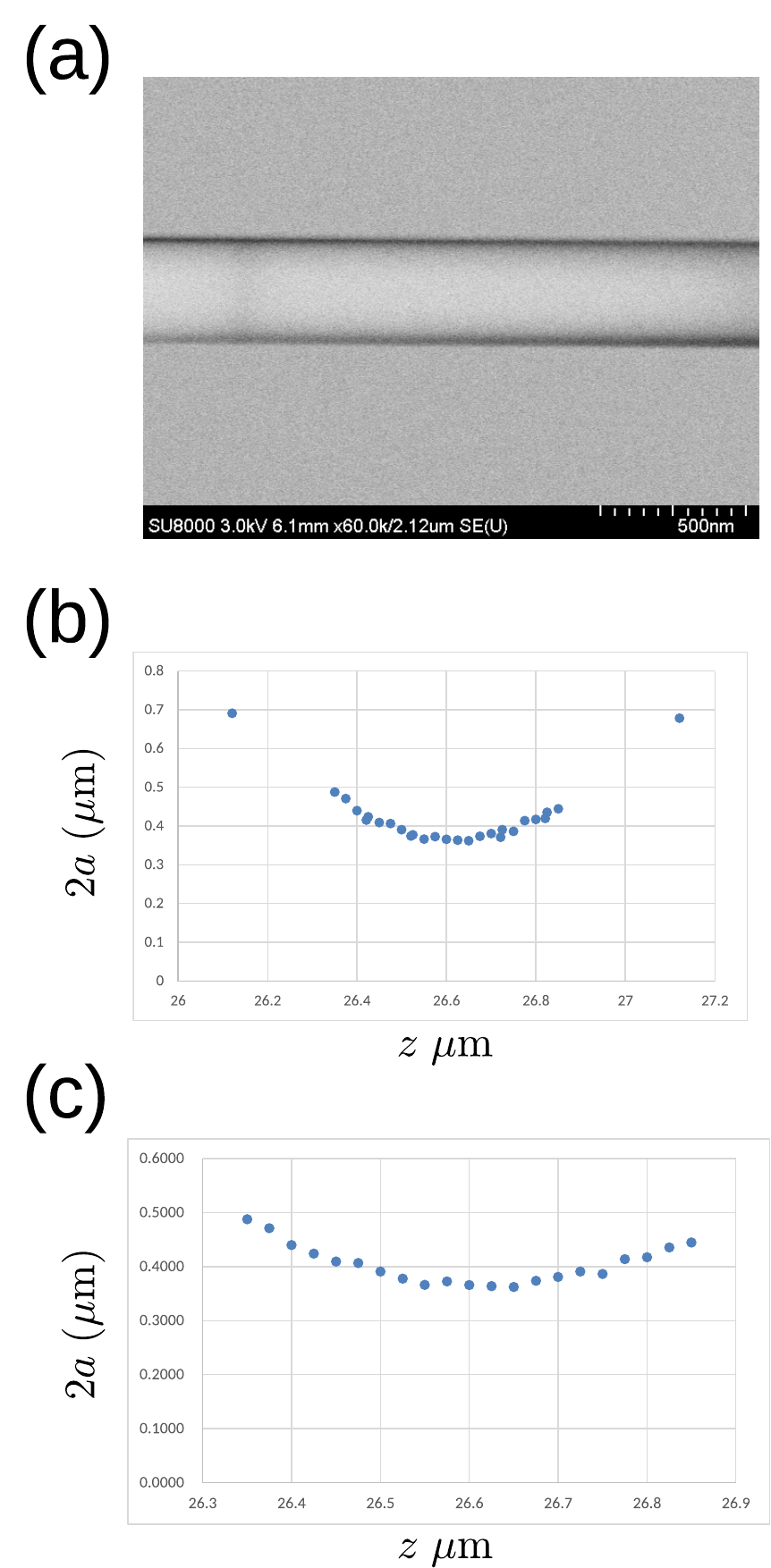}
	\caption[]{(a) Scanning electron microscope of nanofiber used in experiments in its waist region (i.e. sallest diameter region). (b) Diameter measurements of the nanofiber over a $\sim$1 mm range. (c) Diameter measurements over the nanofiber waist region. }
	\label{fig:Nanofiber}
\end{figure}

Fig.~\ref{fig:Nanofiber}(a) shows a scanning electron microscope image taken of the nanofiber within the region of smallest diameter (the ``waist" region). By taking such images along the length of the nanofiber, we can measure its diameter profile. Such a profile over a $\sim$1 mm range is shown in Fig.~\ref{fig:Nanofiber}(b). The nanofiber diameter is sub-micron over this range, and liposome transport along with heating of the water due to absorbance from the fiber mode can in principle happen anywhere along this region.
Fig.~\ref{fig:Nanofiber}(c) shows the sam data as Fig.~\ref{fig:Nanofiber}(b), but zoomed in to the region near the nanofiber waist.

\section*{Liposome size measurements}
Here we display extra diameter characterization data for liposomes extruded through a 100 nm polycarbonate membrane filter. We first show the data presented in the main paper (Fig.~2(a)) to allow easy comparison. Note that the size of the liposomes after filtering can be larger than the filter pore size due to the elasticity of liposomes, whose shape can deform sufficiently to pass through gaps smaller than the nominal liposome diameter.
\begin{figure}[htb]
	\centering
	\includegraphics[width=1.1\linewidth]{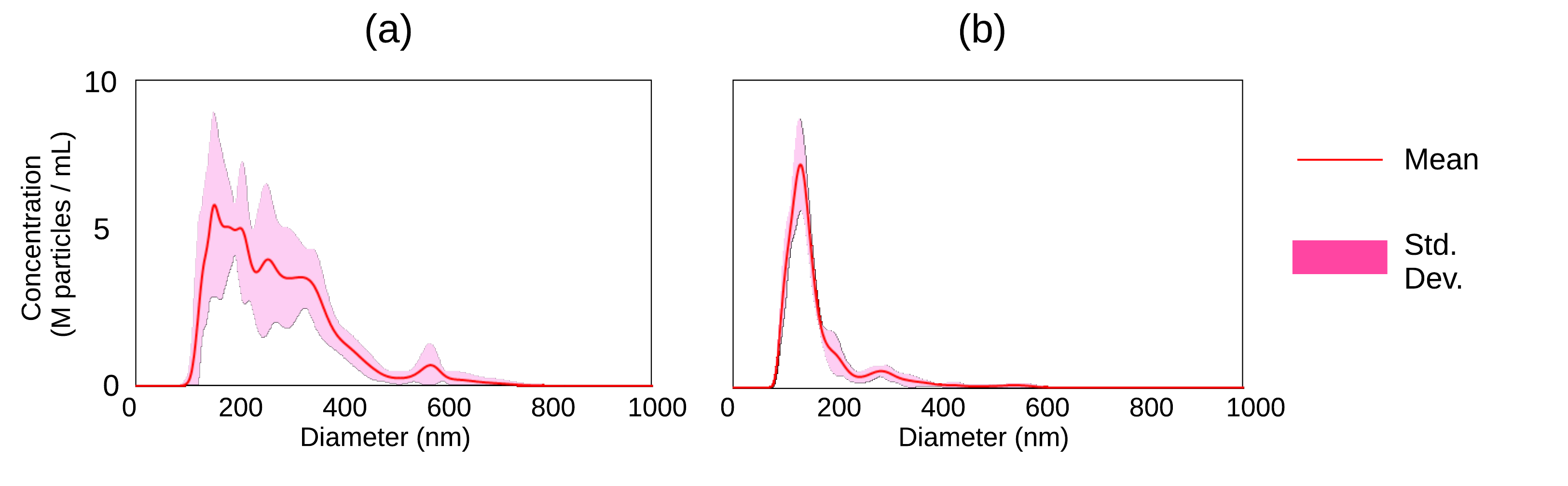}
	\caption[]{(a) Diameter characterization of liposomes used in our experiments. (b) Diameter characterization of liposomes 
	filtered with a 100 nm polycarbonate membrane.}
	\label{fig:Liposome}
\end{figure}

Fig.~\ref{fig:Liposome}(a) shows the diameter characterization data for liposomes filtered through a 400 nm polycarbonate membrane as used in the experiments whose results are given in the main paper. We also attempted experiments with liposomes filtered through a 100 nm polycarbonate membrane (diameter characterization shown in Fig.~\ref{fig:Liposome}(b)) but for that liposome sample, no transport 
behavior was observed, suggesting that transport occurs predominantly for liposomes with diameter greater than 100 nm.
\begin{thebibliography}{99}
\bibitem{Nanobook} Novotny, L., and Hecht, B., ``Principles of nano-optics", Cambridge university press, 2012.
\bibitem{OptMan1} Kawata, S. and Sugiura, T.,``Movement of micrometer-sized particles in the evanescent field of a laser beam", Opt. Lett. \textbf{17}, 772 (1992).
\bibitem{OptMan2} Juan, M., Righini, M., and Quidant, R, ``Plasmon nano-optical tweezers", Nature Photon. \textbf{5}, 349-356 (2011).
\bibitem{SileBubble} Ward, J. M., Yang, Y., Lei, F., Yu, X.-C., Xiao, Y.-F., Nic Chormaic, S. ``Nanoparticle sensing beyond evanescent field interaction with a quasi-droplet microcavity," Optica \textbf{5}, 674-677 (2018).
\bibitem{Aspelmeyer} Magrini, L., Norte, R. A., Riedinger, R., Marinkovi\'{c}, I.,  Grass, D., Deli\'{c}, U., Gr\"{o}blacher,  S., Hong, S. and Aspelmeyer, M., ``Near-field coupling of a levitated nanoparticle to a photonic crystal cavity," Optica \textbf{5}, 1597-1602 (2018).
\bibitem{Ashkin} Ashkin, A., Dziedzic, J. M. and Yamane, T., ``Optical trapping and manipulation of single cells using infrared laser beams", Nature \textbf{330}, 769-771 (1987).
\bibitem{He} He, L., Ozdemir, S. K., Zhu, J., Kim, W., and Yang, L., ``Detecting single viruses and
nanoparticles using whispering gallery microlasers", Nat. Nanotechnol. \textbf{6}, 428–432 (2011).
\bibitem{Beuwer} Beuwer, M. A., Prins, M. W. J., and Zijlstra, P. ``Stochastic protein interactions
monitored by hundreds of single-molecule plasmonic biosensors" Nano Lett. \textbf{15}, 3507–3511 (2015).,
\bibitem{Sandoghdar1} Rattenbacher, D., Shkarin, A., Renger, J., Utikal, T., G\"{o}tzinger, S., and Sandoghdar, V. , ``Coherent coupling of single molecules to on-chip ring resonators", arxiv:1902.05257 (2019).
\bibitem{Vollmer1} Baaske, M. D.,  Foreman, M. R.  and Vollmer, F.,  ``Single-molecule nucleic acid interactions monitored on a label-free microcavity biosensor platform", Nat. Nanotechnol. \textbf{9}, 933 (2014).
\bibitem{Small1} Coles, D., Flatten, L. C.,  Sydney, T., Hounslow, E., Saikin, S. K.,
Aspuru-Guzik, A., Vedral, V., Kuo-Hsiang Tang, J., Taylor, R. A.,Smith, J. M.,  and Lidzey, D. G., ``A Nanophotonic Structure Containing Living Photosynthetic Bacteria", Small \textbf{13}, 1701777 (2017).
\bibitem{Lipson} Yang, A., Moore, S., Schmidt, B. \textit{et al.}, ``Optical manipulation of nanoparticles and biomolecules in sub-wavelength slot waveguides", Nature \textbf{457}, 71-75 (2009).
\bibitem{Gaugiran} Gaugiran, S.,  G\'{e}tin, S., Fedeli, J. M., Colas, G., Fuchs, A., Chatelain, F., D\'{e}rouard, J., ``Optical manipulation of microparticles and cells on silicon nitride waveguides," Opt. Express \textbf{13}, 6956-6963 (2005).
\bibitem{Ahluwalia} Ahluwalia, B. S., McCourt, P., Huser, T., Helleso, O. G., ``Optical trapping and propulsion of red blood cells on waveguide surfaces," Opt. Express \textbf{18}, 21053-21061 (2010).
\bibitem{Helleso} Helleso, O. G., Lovhaugen, P., Subramanian, A. Z., Wilkinson, J. S.,  Ahluwalia, B. S., 
``Surface transport and stable trapping of particles and cells by an optical waveguide loop", Lab Chip, \textbf{12}, 3436-3440 (2012).
\bibitem{Xin} Xin, H., Cheng, C., Li, B., ``Trapping and delivery of Escherichia coli in a microfluidic channel using an optical nanofiber"   Nanoscale, \textbf{5}, 6720-6724 (2013).
\bibitem{Pang} Pang, Y., Gordon, R., ``Optical trapping of a single protein", Nano Lett. \textbf{12}, 402-406 (2012).
\bibitem{Hong} Hong, C., Yang, S., Ndukaife, J.C., ``Stand-off trapping and manipulation of sub-10 nm objects and biomolecules using opto-thermo-electrohydrodynamic tweezers", Nat. Nanotechnol. (2020). \begin{verbatim} https://doi.org/10.1038/s41565-020-0760-z \end{verbatim}
\bibitem{Somasundar} Somasundar, A., Ghosh, S., Mohajerani, F. {\emph et al.}, ``Positive and negative chemotaxis of enzyme-coated liposome motors", Nat. Nanotechnol. \textbf{14}, 1129–1134 (2019).
\bibitem{Osawa} Osawa M., Erickson H.P., ``Liposome division by a simple bacterial division machinery", Proc. Natl. Acad. Sci. USA, \textbf{110} 11000‐11004 (2013).
\bibitem{Mikhaylov} Mikhaylov, G., Mikac, U., Magaeva, A., {\emph et al.}, ``Ferri-liposomes as an MRI-visible drug-delivery system for targeting tumours and their microenvironment", Nat. Nanotechnol. \textbf{6}, 594-602 (2011).
\bibitem{Deng} Deng, W., Chen, W., Clement, S., {\emph et al.}, ``Controlled gene and drug release from a liposomal delivery platform triggered by X-ray radiation", Nat. Commun. \textbf{9}, 2713 (2018).
\bibitem{Fraley} Fraley, R., Dellaporta, S., Papahadjopoulos, D.,  ``Liposome-mediated delivery of tobacco mosaic virus RNA into tobacco protoplasts: A sensitive assay for monitoring liposome-protoplast interactions", Proc. Natl. Acad. Sci. USA \textbf{79}, 1859–1863 (1982).
\bibitem{Godino} Godino, E., L\'{o}pez, J.N., Foschepoth, D., {\emph et al.}, ``De novo synthesized Min proteins drive oscillatory liposome deformation and regulate FtsA-FtsZ cytoskeletal patterns", Nat. Commun. \textbf{10}, 4969 (2019).
\bibitem{SizePaper} Bendix, P. M.  and Oddershede, L. B., ``Expanding the Optical Trapping Range of Lipid Vesicles to the
Nanoscale", Nanoletters \textbf{11}, 5431 (2011).
\bibitem{Spyratou} Spyratou, E., Cunaj, E., Tsigaridas, G., Mourelatou, E. A., Demetzos, C., Serafetinides, A. A., Makropoulou, M., ``Measurements of liposome biomechanical properties by combining line optical tweezers and dielectrophoresis", Journal of liposome research, \textbf{25}, 202-210. (2015)
\bibitem{Exo1} Im, H., Shao, H., Park, Y., Peterson, V. M., Castro, C. M., Weissleder, R., and Lee Hakho, ``Label-free detection and molecular profiling of exosomes with a nano-plasmonic sensor", Nature Biotechnology \textbf{32}, 490-495 (2014). 
\bibitem{Exo2} Ndao, A., Hsu, L., Cai, W., Ha, J., Park, J., Contractor, R., Lo, Y., and Kant\'{e}, B. (2020). ``Differentiating and quantifying exosome secretion from a single cell using quasi-bound states in the continuum", Nanophotonics,  20200008 (2020).
\bibitem{thermal} Hill, E. H., \textit{et. al.} ``Opto-Thermophoretic Attraction, Trapping, and Dynamic Manipulation of Lipid Vesicles", Langmuir \textbf{34}, 13252-13262 (2018).
\bibitem{nanoplasmon} Kang, Z., Chen, J., Wu, S.-Y., Chen, K., Kong, S.-K., Yong, K.-T, and Ho H.-P., ``Trapping ensembles of particles and live cells on large-scale random gold nano-island substrates", Scientific Reports \textbf{5}, 9978 (2015).
\bibitem{kaneta1} Kuboi M., Takeyasu, N., Kaneta T., ``Enhanced Optical Collection of Micro- and Nanovesicles in the Presence of Gold Nanoparticles", ACS Omega \textbf{3}, 2527-2531 (2018).
\bibitem{kaneta2} Tani, Y., Kaneta, T., ``Enhancement of optical force acting on vescicles via the binding of gold nanoparticles", Royal Society Open Science \textbf{6}, 190293 (2019).
\bibitem{Skelton} Skelton, S. E. \textit{et al.}, ``Evanescent wave optical trapping and transport of micro- and nanoparticles on tapered optical fibers", J. Quant. Spectrosc. Radiat. Transf. 113, 2512–2520 (2012).
\bibitem{Sile} Maimaiti, A., \textit{et al.} ``Higher order microfibre modes for dielectric particle trapping and propulsion", Sci. Rep. \textbf{5}, 9077 (2015).
\bibitem{HiranoLab1} Yamaura, D., \textit{et al.}, ``Amphiphobic Septa Enhance the Mechanical Stability of Free-Standing Bilayer Lipid Membranes", Langmuir \textbf{34}, 5615-5622 (2018).
\bibitem{Nele} Nele, V., Holme, M. N., Kauscher, U., Thomas, M. R., Doutchm, J. J., and Stevens, M. M., ``Effect of formulation method, lipid composition and PEGylation on Vesicle lamellarity: A small angle neutron scattering study", Langmuir \textbf{35}, 6064 (2019).
\bibitem{Goldman}
Goldman, A.J., Cox, R.G., Brenner, H., ``Slow motion of a sphere parallel to a plane wall - I Motion through a quiescent fluid",
Chemical Engineering Science \textbf{22}, 637-651 (1967).
\bibitem{Alnayli} Alnayli, R., Shanon, S., and Hadi, A., ``Study the Linear and Nonlinear Optical Properties for Laser Dye Rhodamine B",
J. Phys.: Conf. Series \textbf{1234} 012022, (2019).
\bibitem{Matsuzaki} Matsuzaki,   \textit{et al.}, ``Optical characterization of liposomes by right angle light scattering and turbidity measurement", Biochimica et Biophysica Acta \textbf{1467} 219-226 (2000).
\bibitem{SileHP} Ward, J. M., Maimaiti, A., Le, V. H., and Nic Chormaic, S., ``Optical micro- and nanofiber pulling rig" Rev. Sci.
Instrum. \textbf{85} 111501 (2014).
\bibitem{ChandraQED} Yalla, R., Sadgrove, M.,  Nayak, K. P., and Hakuta, K., ``Cavity quantum electrodynamics on a nanofiber using a composite photonic crystal cavity", Phys. Rev. Lett. \textbf{113}, 143601 (2014).
\bibitem{KaliReview} Nayak, K. P., {\emph et al.}, ``Nanofiber quantum photonics", J. Opt. \textbf{20} 073001 (2018).
\bibitem{BirksLi} Birks, Timothy A., and Li, Y. W., ``The shape of fiber tapers." Journal of Lightwave Technology \textbf{10} 432-438 (1992).
\bibitem{MacDonald} MacDonald, R. C, MacDonald, R. I., Menco, B. P. M., Takeshita, K., Subbarao, N. K., Hu, L. R., ``Small-Volume Extrusion Apparatus for Preparation of Large, Unilamellar Vesicles", Biochim. Biophys. Acta - Biomembr. \textbf{1061}, 297–303 (1991).
\bibitem{Lumerical} Lumerical Inc. ``Methodology for optical force calculations",\\
\begin{verbatim}https://support.lumerical.com/hc/en-us/articles/360042214594\end{verbatim}
\end{thebibliography}
\end{document}